\documentclass[a4paper,11pt]{article}
\usepackage{graphicx}
\usepackage{cite}

\def \d {{\rm d}}

\def \U {{\cal U}}
\def \V {{\cal V}}

\usepackage{a4wide}
\usepackage{amsthm}
\usepackage{amsmath}
\usepackage{amssymb}
\usepackage{caption}

\newcommand{\Up}{U_+}

\newtheorem{thm}{Theorem}[section]
\newtheorem{lem}[thm]{Lemma}
\newtheorem{cor}[thm]{Corollary}
\newtheorem{rem}[thm]{Remark}

\title{The global uniqueness and $C^1$-regularity of geodesics in expanding impulsive gravitational waves}
\author{
J.~Podolsk\'y$^1$\thanks{{\tt podolsky@mbox.troja.mff.cuni.cz}},
C.~S\"amann$^2$\thanks{{\tt clemens.saemann@univie.ac.at}},
R.~Steinbauer$^2$\thanks{{\tt roland.steinbauer@univie.ac.at}}
and R.~\v{S}varc$^1$\thanks{{\tt robert.svarc@mff.cuni.cz}} \\ \\
$^1$ Institute of Theoretical Physics,\\
Charles University in Prague, Faculty of Mathematics and Physics, \\
V Hole\v{s}ovi\v{c}k\'ach 2, 18000 Prague 8, Czech Republic.\\ \\
$^2$ Faculty of Mathematics, University of Vienna, \\
Oskar-Morgenstern-Platz 1, 1090 Vienna, Austria. \\ \\
}
\date{July 28, 2016}

\begin{document}

\maketitle

\begin{abstract}
We study geodesics in the complete family of expanding impulsive
gravitational waves propagating in spaces of constant curvature, that is
Minkowski, de~Sitter and anti-de~Sitter universes.
Employing the continuous form of the metric we rigorously prove existence and
global uniqueness of continuously differentiable geodesics (in the sense of
Filippov) and study their interaction with the impulsive wave. Thereby we
justify the ``$C^1$-matching procedure'' used in the literature to
derive their explicit form.
\end{abstract}

\section{Introduction}
Impulsive gravitational waves for some time now have served as simple yet
interesting models of exact radiative spacetimes in Einstein's theory describing violent but
short bursts of gravitational radiation, see e.g.\ \cite[Ch.\ 20]{GP:09}. Also
they are spacetimes of low regularity described either by a (locally Lipschitz)
continuous metric or even by a distributional metric. Consequently, these
geometries are also interesting from a mathematical point of view, raising
questions in non-smooth Lorentzian geometry --- a topic that has recently
attracted some attention (e.g.\ \cite{CG:12, Min:15, Sbi:15, Sae:15, KSSV:15}).

Indeed, in the case of impulsive \emph{pp}-waves \cite[Sec.\ 20.2]{GP:09}, i.e.,
\emph{nonexpanding} impulsive waves in \emph{Minkowski space}, the
discontinuous transformation between the distributional Brinkman form of the
metric and the Lipschitz continuous Rosen form has been put into the
mathematically rigorous framework of nonlinear distributional geometry in
\cite{KS:99}. At the heart of this result lies a good mathematical understanding
of the geodesics in both forms of the metric. With the long-term objective in
mind to generalise this result to nonexpanding impulsive waves propagating on
\emph{all backgrounds of constant curvature} with any cosmological constant~$\Lambda$,
recently their geodesics have been studied in the continuous form \cite{PSSS2015} as well as in the distributional form \cite{SSLP2015} (using a 5D-formalism). In particular, in
the continuous form it was essential to use a general solution concept due to
Filippov \cite{F:88} --- well known in ODE-theory --- to cope with the geodesic
equation which has a discontinuous but bounded right hand side.

In this work we transfer this approach to \emph{expanding} impulsive waves, see
e.g. \cite[Sec. 20.4--5]{GP:09}. More precisely, we consider  the \emph{entire
class of expanding impulsive waves}
propagating on \emph{spaces of constant curvature} --- Minkowski space, de Sitter
and anti-de Sitter universes (with vanishing, positive and negative cosmological
constant $\Lambda$, respectively). It is well known that the mathematical
intricacies connected with the distributional form of the metric and its
relation to the continuous form are much more severe in the expanding case.
Nevertheless relevant progress has been achieved in
\cite{[B8],[B10],GriffithsDochertyPodolsky:2004,PodolskyGriffiths2004} --- although
partly only formal. On the other hand, using the continuous form of the metric
the geodesics have been explicitly described in \cite{PodolskySteinbauer2003}
for Minkowski background and in \cite{PodolskySvarc2010} for general $\Lambda$.
Both of these works used a ``$C^1$-matching procedure'': The geodesics of the
background spacetime on both ``sides'' of the impulsive wave were matched on the
wave surface. However, to obtain the correct number of equations to match all
integration constants ``before'' and ``behind'' the wave impulse it had to be
\emph{assumed --- without proof --- }that the geodesics are continuously
differentiable curves. It is the main objective of this work to supply such a
proof.

We begin, however, in Section \ref{sec:eigw} with a rather detailed review of
the complete class of expanding impulsive gravitational waves in spaces of constant
curvature, including various methods of their construction. In particular, we
collect all the main forms of the metric in a unified notation to also provide
a point of reference for future work. We focus on particle motion using the
continuous form of the metric in Section \ref{sec:geo}. We briefly review
previous work \cite{PodolskySteinbauer2003,PodolskySvarc2010} and derive the
equations for the real form of the metric in Section~\ref{ssec:geo}. Then,
in Section~\ref{sec:ex} we employ the \emph{Lipschitz property} of the
continuous form of the metric which allows for an application of
Filippov's solution theory for ordinary differential equations with
discontinuous right hand sides to solve the geodesic equations. In this way the
\emph{existence} and the \emph{$C^1$-regularity} of the geodesics is obtained from a
general result \cite{S:14}. However, the quest for uniqueness becomes
delicate since it is no longer possible to argue on general grounds
(cf.~\cite[Sec.\ 3.3]{PSSS2015}) but we have to combine arguments exploiting the
geometry of the spacetimes at hand with basic facts from Filippov's theory. In
particular, we provide a detailed study of the interaction of the geodesics
with the wave impulse and in this way we prove in Section \ref{sec:uni} that
the geodesic equations possess \emph{globally unique} continuously
differentiable solutions. This turns the ``$C^1$-matching procedure''\!, employed
in \cite{PodolskySteinbauer2003,PodolskySvarc2010} and reviewed in Section
\ref{sec:explgeod}, into a mathematically valid technique to explicitly derive
the geodesics that \emph{cross} the impulse. Moreover, we also find
(spacelike) geodesics that \emph{touch} the impulse which have not been 
considered in the context of the matching procedure so far.

\section{Exact expanding impulsive gravitational waves in spacetimes of
constant curvature}\label{sec:eigw}

Physically, impulsive gravitational waves arise most naturally as a limit of a
suitable family of sandwich waves with profiles of ever ``shorter duration''
$\varepsilon$ which simultaneously become ``stronger'' as ${\varepsilon^{-1}}$.
Mathematically, this amounts to a distributional limit of a sequence of sandwich
profiles which converges to the profile $\delta$, the Dirac function. An
impulsive gravitational wave is thus localised on a \emph{single wave-front},
which is a null hypersurface.

Interestingly, there exist several alternative methods of construction of such
exact expanding solutions to Einstein's vacuum field equations. They will now be
summarised and compared, together with the appropriate references to
original works.

\subsection{The Penrose ``cut and paste'' method}

A fundamental geometric method for constructing impulsive (purely) gravitational
spherical waves, expanding in backgrounds of constant curvature, was introduced
(for flat space) by Penrose in his seminal work \cite{Pen72}. The general method
starts with the following unified form of Minkowski (${\Lambda=0}$) or
(anti-)de~Sitter
(${\Lambda\not=0}$) spacetime
 \begin{equation}
\d s_0^2= \frac{2\,\d\eta\,\d\bar\eta -2\,\d\U\,\d\V}
{[\,1+\frac{1}{6}\Lambda(\eta\bar\eta-\U\V)\,]^2}\,,
 \label{conf}
 \end{equation}
on which the transformation
\begin{equation}
  \eta = \frac{Z}{p}\,V\,, \quad
  \U   = \frac{Z\bar Z}{p}\,V-U \,, \quad
  \V   = \frac{1}{p}\,V-\epsilon U
  \label{inv}
%\end{equation}
\quad\text{with}\quad
%\begin{equation}
  p=1+\epsilon Z\bar Z\,, \quad \epsilon=-1,0,+1
%  \label{pandep}
\end{equation}
is applied. The background spacetimes of constant curvature thus take the form
  \begin{equation}
\d s_0^2 = \frac{ 2\,(V/p)^2\,\d Z\,\d\bar Z +2\,\d U\,\d V -2\epsilon\,\d U^2}
{\big[\,1+\frac{1}{6}\Lambda U(V-\epsilon U)\,\big]^2}\,.
  \label{U<0}
  \end{equation}
In these coordinates, the hypersurface  ${U=0}$ is a \emph{future null
cone}~$\cal N$ (a sphere expanding with the speed of light) since ${U(V-\epsilon U)=\eta\bar\eta-\U\V = 0}$. The Minkowski
or (anti-)de~Sitter manifold ${\cal M}$ can thus be divided into two parts,
namely ${{\cal M}^-(U<0)}$ inside the null cone~$\cal N$, and  ${{\cal
M}^+(U>0)}$ outside of it.

The Penrose ``cut and paste'' construction is based on
re-attaching these two parts ${{\cal M}^-}$ and ${{\cal M}^+}$ with a
particular ``warp'' along~$\cal N$, generated by an \emph{arbitrary complex
valued function} $h(Z)$, see Figure~\ref{conicalcut}.
Specifically, the \emph{Penrose junction conditions} prescribe the identification
\begin{equation}
\Big[\,Z,\,\bar Z,\,V,\,U=0_-\,\Big]_{_{{\cal M}^-}}\equiv
\Big[\,h(Z),\,\bar h({\bar Z}),\,
\frac{1+\epsilon\, h\bar h}{1+\epsilon Z\bar Z} \frac{V}{|h'|},
\,U=0_+\,\Big]_{_{{\cal M}^+}}
\  \label{juncexp}
\end{equation}
of the corresponding points from the two re-attached parts across the expanding
sphere ${U=0}$.

\begin{minipage}{.38\textwidth}
%\begin{figure}[htp]
%\centerline{\includegraphics[scale=0.5]{conicalcut} }
\vspace*{6pt}
\centering{\includegraphics[scale=0.4]{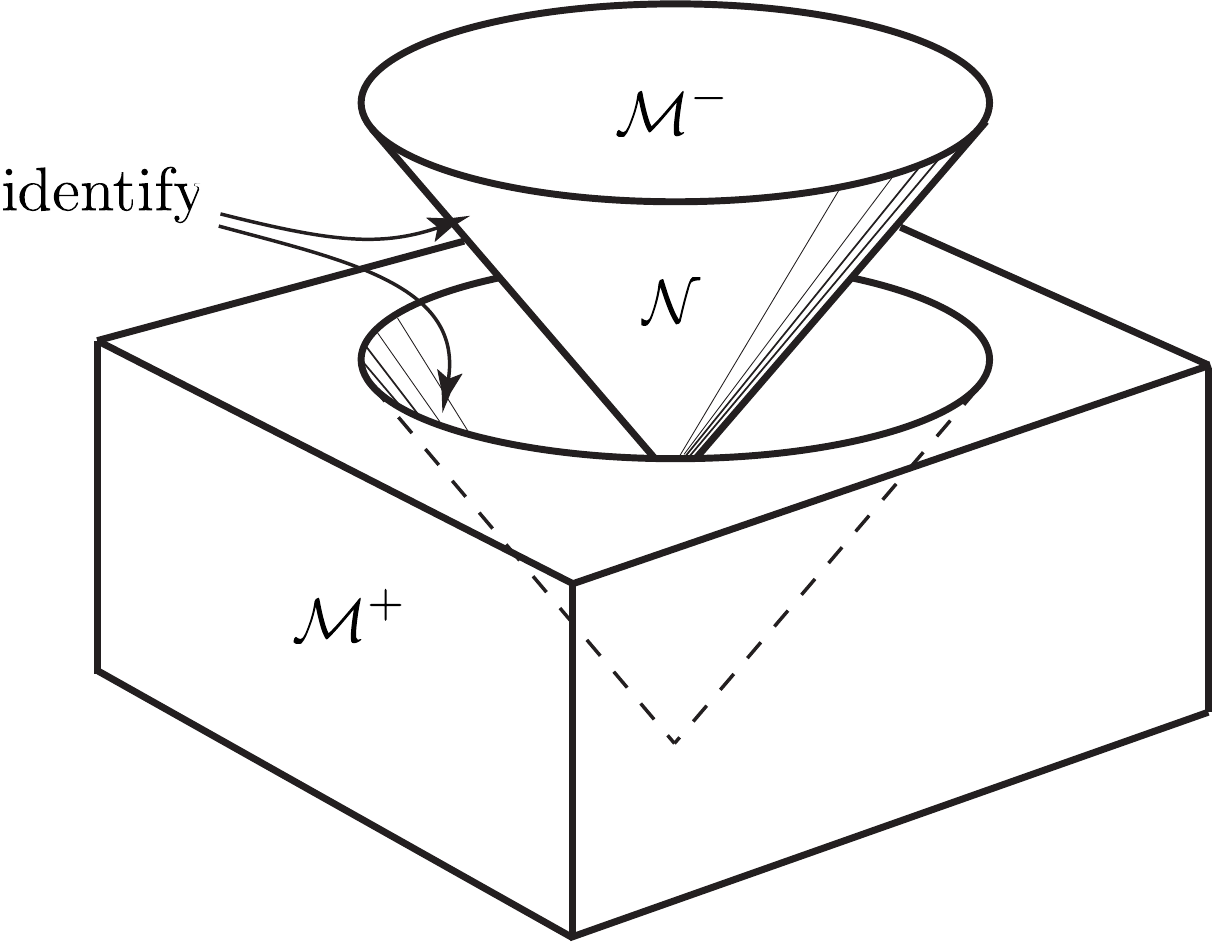} }
%\end{figure}
\vspace*{6pt}
\end{minipage}
\hfill
\begin{minipage}{.6\textwidth}
\captionof{figure}{ \small Minkowski or (anti-)de Sitter space is cut into two
parts ${\cal M}^-$ and ${\cal M}^+$ along a future null cone~${\cal N}$. These
parts are then re-attached with an arbitrary ``warp'' in which points on both
sides of~${\cal N}$ are identified. Such a construction generates spherical
impulsive gravitational waves expanding in these constant-curvature
backgrounds.}
\label{conicalcut}
\end{minipage}

In \cite{Pen72} Penrose only considered the
%introduced this construction for the simplest
case ${\Lambda=0}$, ${\epsilon=0}$, see also
\cite{NutPen92,Nutku}.\footnote{A similar (yet
different) ``cut and paste'' construction was employed by Gleiser and Pullin
\cite{GlePul89} to obtain a specific solution,
namely a spherical impulse generated by a ``snapping'' cosmic string in flat
space%, see the metric (\ref{GP}), (\ref{ABsnap})
, see also Section \ref{sec:2.4}.}
The generalisation to the cases ${\Lambda\not=0}$, ${\epsilon=0}$ and
${\Lambda=0}$, ${\epsilon=+1}$ was found by Hogan in \cite{Hogan92} and
\cite{Hogan94}, respectively (see also \cite{Nutku} for arbitrary $\epsilon$).
The completely general form (\ref{U<0}), (\ref{juncexp})
%of the Penrose junction conditions for both arbitrary $\Lambda$ and $\epsilon$
was subsequently found by Podolsk\'y and Griffiths \cite{[B8], [B9], [B10]}.

\subsection{Continuous coordinates}

The Penrose ``cut and paste'' method, although illustrative, does
not provide explicit metric forms of the complete spacetimes. We now do so,
following and extending Hogan \cite{Hogan93,Hogan94}, and perform another
transformation of (\ref{conf}), generalising (\ref{inv}) but still linear in $U$
and $V$, given by
\begin{align}\label{transe3}
\V   = {\cal A}\,V-{\cal D}\,U\,, \quad
\U   = {\cal B}\,V-\,{\cal E}\,U\,, \quad
\eta\, = \,{\cal C}\,V-{\cal F}\,U\,,
 \end{align}
where
 \begin{eqnarray}
&&{\cal A}= \frac{1}{p|h'|}\,,\quad
{\cal B}= \frac{|h|^2}{p|h'|}\,,\quad
{\cal C}= \frac{h}{ p|h'|}\,,  \quad  %\nonumber\\
%&&
{\cal D}= \frac{1}{|h'|}\bigg\{
\frac{p}{4} \left|\frac{h''}{h'}\right|^2+\epsilon
\left[1+\frac{Z}{2}\frac{h''}{h'}+\frac{\bar Z}{2}\frac{\bar h''}{\bar h'}
\right]\!\bigg\}, \nonumber \\
&&{\cal E}= \frac{|h|^2}{|h'|}
\bigg\{ \frac{p}{4}\left|\frac{h''}{h'}-2\frac{h'}{h}\right|^2
+\epsilon\left[ 1+\frac{Z}{2}
\left(\frac{h''}{h'}-2\frac{h'}{h}\right)+\frac{\bar Z}{2}
\left(\frac{\bar h''}{\bar h'}-2\frac{\bar h'}{\bar h}\right)
\right]\!\bigg\}\,,\label{transe4}\\
&&{\cal F}= \frac{h}{|h'|}\bigg\{
\frac{p}{4}\left(\frac{h''}{h'}-2\frac{h'}{h}\right)
\frac{\bar h''}{\bar h'}
+\epsilon\left[1+
 \frac{Z}{2}\left(\frac{h''}{h'}-2\frac{h'}{h}\right)
+\frac{\bar Z}{2}\frac{\bar h''}{\bar h'}\right]\!\bigg\}\,,\nonumber
 \end{eqnarray}
and ${h=h(Z)}$ is as above. %arbitrary complex function.
%, and the derivative with respect to its argument $Z$ is denoted by a prime.
Interestingly, the coefficients (\ref{transe4}) satisfy the non-trivial
identities
 \begin{equation}
{\cal C}\bar {\cal C}-{\cal A}{\cal B}=0\,,\quad
{\cal F}\bar {\cal F}-{\cal D}{\cal E}=-\epsilon\,,\quad
{\cal A}{\cal E}+{\cal B}{\cal D}-{\cal C}\bar {\cal F}-\bar {\cal C}{\cal
F}=1\,,
 \label{identitiesABCDEF}
 \end{equation}
implying ${\eta\bar\eta-\U\V = U(V-\epsilon U)}$. The null cone $\cal N$ is thus
again located along ${U=0}$.

With the transformation (\ref{transe3}), (\ref{transe4}), the  metric
(\ref{conf}) of any constant curvature space becomes
 \begin{equation}
 \d s_0^2 = \frac{2\left| (V/p)\d Z+Up\bar H\d\bar Z \right|^2
  +2\d U\d V -2\epsilon\d U^2}
{\big[\,1+\frac{1}{6}\Lambda U(V-\epsilon U)\,\big]^2}\,,
 \label{U>0}
 %\end{equation}
 \ \text{with}\
 %\begin{equation}
 H(Z)=\frac{1}{2}\!
\bigg[\frac{h'''}{h'}-\frac{3}{2}\left(\frac{h''}{h'}\right)^{\!2}
\bigg].  %\label{Schwarz}
 \end{equation}
Notice that (\ref{transe3}), (\ref{transe4}) reduce to (\ref{inv}) for the
simplest choice ${h(Z)=Z}$ implying ${H=0}$.
We now combine the line element (\ref{U>0}) for ${U>0}$ with
%and attach this to
the metric (\ref{U<0}) for ${U<0}$ to obtain
%, which was obtained from (\ref{conf}) by the transformation (\ref{inv})
%corresponding to ${h(Z)=Z}$. The resulting full line
%element thus takes the form
 \begin{equation}
\d s^2 =  \frac{2\left| (V/p)\,\d Z+U_+(U)\,p\,\bar H\,\d\bar Z \right|^2+2\,\d
U\,\d V -2\epsilon\,\d U^2}
{\big[\,1+\frac{1}{6}\Lambda U(V-\epsilon U)\,\big]^2}\,,
 \label{en0}
 \end{equation}
where ${U_+(U)}$ is the \emph{kink function} defined as
${U_+(U)\equiv 0}$ for ${U\le0}$ and ${U_+(U)\equiv U}$ for ${U\ge0}$, i.e., ${U_+=U
\Theta(U)}$ where $\Theta$ is the Heaviside step function. This metric was
presented for ${\Lambda=0}$ in \cite{NutPen92,Hort:1990,Hogan93,Hogan94},
for ${\Lambda\not=0}$ in \cite{Hogan92}, and in the most
general form in \cite{[B8],[B10]}.\footnote{Another continuous metric
generalising (\ref{en0}) for ${\Lambda=0}$ was found in \cite{NutAli},
extending results for spherical shock waves \cite{Nutku}. It contains
an additional parameter related to acceleration of the coordinate system.}

Since the kink function is Lipschitz continuous the metric (\ref{en0}) is locally
Lipschitz in the variable $U$. Thus, apart from possible singularities of the function $H$, the spacetime is locally Lipschitz. Recall that by Rademacher's theorem a locally Lipschitz metric $g$ (denoted by ${g\in C^{0,1}}$) possesses a locally bounded connection, and so the metric is well within the ``maximal'' distributional curvature framework as identified by Geroch and Traschen \cite{GT:87}: Indeed  a metric of
Sobolev regularity $H^{1}_{\mbox{\small loc}}\cap L^\infty_{\mbox{\small loc}}$ allows to (stably) define the curvature in distributions, see also \cite{LM:07,S:08}.
Since locally Lipschitz metrics possess no bound on the curvature (in $L^\infty$),
the discontinuity in the derivatives of the metric introduces impulsive components in the Weyl and curvature tensors, namely ${\Psi_4=(p^2H/ V)\,\delta(U)}$ and
${\Phi_{22}=(p^4H\bar H/ V^2)\,U\,\delta(U)}$.
Clearly, the spacetime is thus conformally flat everywhere except on the
impulsive wave surface ${U=0}$.
It is a vacuum solution everywhere except at ${V=0}$ on the wave surface ${U=0}$
where a curvature singularity (``origin of the impulse'') is located, and at
possible poles of the function $p^2H$. Hence it
is most natural to only consider the region ${V>0}$ of the spacetime.

The above procedure is an \emph{explicit version} of the ``cut and paste''
construction since by comparing the transformations (\ref{inv}) at ${U=0_-}$ with
(\ref{transe3}), (\ref{transe4}) at ${U=0_+}$, we obtain exactly (\ref{juncexp}).
\smallskip

\noindent
{\bf The geometrical meaning of the function $h(Z)$. }\label{geometricalmeaning}
The generating complex function ${h(Z)}$ provides a geometric interpretation of
the junction conditions (\ref{juncexp}), see \cite{NutPen92,[B10]}:
Evaluating the ratio ${\eta/\V}$ using (\ref{inv}) and (\ref{transe3})
for ${U<0}$ and ${U>0}$, respectively, we find that \emph{on the impulse}
 \begin{equation}
\frac{\eta}{\V}=Z    \quad  \mbox{for} \quad U=0_-\,,
\qquad \hbox{and}\qquad
\frac{\eta}{\V}=h(Z) \quad  \mbox{for} \quad U=0_+\,.
 \label{xi}
 \end{equation}
By (\ref{conf}) we have ${\eta/\V=(x+{\hbox{i}}\,y)/(t-z)}$ in
Minkowski  %(and ${\eta/\V=(Z_2+{\hbox{i}}\,Z_3)/(Z_0-Z_1)}$ in
and also (anti-)de~Sitter space,
see \cite[Ch. 4--5]{GP:09} or \cite{PodolskySvarc2010}.
This is the relation for a \emph{stereographic projection}
from the North pole of the sphere  onto its equatorial plane. This
permits us to represent the wave surface ${U=0}$  either as a Riemann sphere
or as its associated complex plane parametrized by the coordinate $Z$.
Accordingly, the Penrose junction condition (\ref{juncexp})
can equivalently be understood as a \emph{mapping on the complex plane}
${Z\to h(Z)}$.

This insight can be used to construct explicit solutions: For example, we may
assume that the region
${U<0}$ \emph{inside} the impulse represented
by ${Z=|Z|e^{i\phi}}$  covers the \emph{complete} sphere, ${\phi\in[-\pi,\pi)}$.
However, the range of the function $h(Z)$ in general will not cover the
\emph{entire} sphere
outside the spherical impulse for $U>0$. In particular, the complex mapping
 \begin{equation}
 h(Z)=Z^{1-\delta}\,,
 \label{h1}
 \end{equation}
where ${\delta>0}$, covers the plane minus a wedge as  ${\,\arg
h(Z)\in[-(1-\delta)\pi,(1-\delta)\pi)}$. This represents Minkowski, de Sitter, or anti-de~Sitter  space with a
\emph{deficit angle} $2\pi\delta$, which may be considered to \emph{describe a snapped cosmic string}
in the region \emph{outside} the spherical impulsive wave. The string has a
constant tension and is located along the axis ${\eta=0}$. The corresponding metric takes the form (\ref{en0}) with $H$ generated from (\ref{h1}), i.e.
$% \begin{equation}
 H={\textstyle\frac{1}{2}\delta(1-\frac{1}{2}\delta)}\,Z^{-2}\,,
 %\label{H1}
 %\end{equation}
$
see \cite{[B10]} for more details.
% Let us also remark that
Also quantum fluctuations and aspects of particle creation
on such expanding spherical impulsive and shock waves were analysed (in
different coordinates) by Horta\c csu \cite{Hort:1990, Hort:1992} and his
collaborators \cite{Hort:1995,Hort:1996a,Hort:1996b}.
More generally one may, e.g.,  construct impulsive waves
generated by  \emph{two colliding and snapping cosmic strings} \cite{NutPen92},
see also \cite{[B10],Kar:15}.
\smallskip

\noindent
{\bf Contracting and expanding impulses.} Hogan in \cite{Hogan95} has considered a
natural extension in which the impulse in addition to the future null cone~${\cal N}$
is also located along the \emph{past null cone}. Such a
spacetime contains \emph{both imploding and exploding impulses}, with a
curvature singularity at the common vertex. We now extend Hogan's construction
\cite{Hogan95}  to arbitrary $\Lambda$ and~${\epsilon}$
by introducing ${V'=V-\epsilon U}$ and modifying (\ref{en0}) to
 \begin{equation}
\d s^2 =  \frac{2\left| (V'+\epsilon U)/p\,\d Z+\Big(U_+(U)\,\bar
H+[V'-V_+(V')]\,\bar G\Big)\,p\,\d\bar Z \right|^2
+2\,\d U\,\d V'}
{\big(\,1+\frac{1}{6}\Lambda UV'\,\big)^2}\,.
 \label{en0implexpl}
 \end{equation}
Here the two complex functions $H(Z)$ and $G(Z)$ characterise
the expanding and the contracting impulse, respectively.
The complete null cone is now given by ${\eta\bar\eta-\U\V = UV'=0}$, and the
Weyl tensor components are ${\Psi_4=(p^2H/ V')\,\delta(U)}$ and
${\Psi_0=-\epsilon (p^2\bar G/U)\,\delta(V')}$, with $\Psi_4$ and $\Psi_0$
representing the exploding and the imploding impulse, respectively.
Such a spacetime is algebraically general. At ${U=0=V'}$ there is a highly complicated physical singularity.

\subsection{Limits of sandwich waves}

We now turn to the construction of expanding impulsive waves as distributional limits of sandwich waves in a suitable family of exact radiative spacetimes --- as mentioned at the beginning of this section. It was explicitly argued in \cite{[B8]} that the full family of solutions for expanding spherical gravitational waves can be considered to be an impulsive limit of the class of \emph{vacuum Robinson--Trautman type~N
solutions} with a cosmological constant.
\smallskip

\noindent
{\bf Robinson--Trautman sandwich waves. }
The standard metric \cite{RT1960, RT1962} of Robinson and Trautman (see also
\cite{NewmanTamburino:1962} and \cite{KSMH,GP:09}\footnote{To achieve a
consistency throughout this review we introduced an inversion ${u\to-u}$ as
compared to \cite{GP:09}. Also notice a different scaling gauge
${\zeta\to\sqrt{2}\zeta}$, ${u\to\sqrt{2}u}$, ${r\to r/\sqrt{2}}$ which gives the factor 2 in the term $2\epsilon$.}) reads
 \begin{equation}
 \d s^2 = 2\frac{r^2}{P^2}\,\d\zeta\d\bar\zeta + \,2\,\d u\,\d r
 -\Big(2\epsilon +2r(\log P)_{,u}-{\textstyle\frac{1}{3}}\Lambda r^2\Big)\d u^2
\,,
 \label{RTNmetric}
 \end{equation}
in which the function $P(\zeta,\bar\zeta,u)$ has the general form
\cite{FosterNewman:1967}
 \begin{equation}
 P=\left(1+\epsilon  F\bar F\right)\big(F_{,\zeta}\bar
F_{,\bar\zeta}\big)^{-1/2},
 \label{RTP}
 \end{equation}
where ${F=F(\zeta,u)}$ is an \emph{arbitrary} complex valued function of $u$,
holomorphic in $\zeta$, and ${\epsilon =-1,0,+1\,}$  determines the
Gaussian curvature ${K=2\epsilon}$ of each wave surface ${u=\>}$const. on
${r=1}$, spanned by~$\zeta$.
For the simplest case ${F=\zeta}$ and ${\Lambda=0}$ we obtain the metric
(\ref{U<0}) of Minkowski space, with the identification ${u=U}$, ${r=V}$,
${\zeta=Z}$, and ${P=p}$.

As shown in \cite{RT1960, RT1962} and the work by Newman and Unti
\cite{NewmanUnti:1963}, recently reviewed and generalised in
\cite{Podolsky:2011}, the coordinates employed in (\ref{RTNmetric}) are the most
natural ones for twist-free spacetimes, having a clear geometrical meaning:
Consider \emph{any worldline} $\gamma$ in flat space. At any event ${\cal P}$ on
$\gamma$ \emph{construct the future null cone} $C_\tau=\{u=\tau\}$, where $\tau$ is the parameter value of ${\cal P}$ along $\gamma$. The resulting family of null cones
(locally) foliates the spacetime.
Now, introduce the coordinate $r$ as the \emph{affine parameter} along the
null generators of $C_\tau$, normalised such that ${r=0}$ labels
${\cal P}\in\gamma$.\footnote{In fact, the
vector field $\partial_r$ generates the (quadruply degenerate) principal null
congruence which is geodesic, shear-free, twist-free and expanding in the case
of metric \eqref{RTNmetric}.} Finally,
introduce \emph{two spatial coordinates}
to label all points on the sections ${r =\>}$const.\ on $C_{\tau}$. In the case
${\epsilon=+1}$ this is a 2-sphere, most naturally paramete\-rized by
${\theta\in[0,\pi]}$, ${\phi\in[-\pi,\pi)}$ via
%, or equivalently the complex stereographic coordinate
${\zeta\equiv\tan(\theta/2)\exp(\hbox{i}\phi)}$.

\begin{figure}[htp]
\centerline{\includegraphics[scale=0.4]{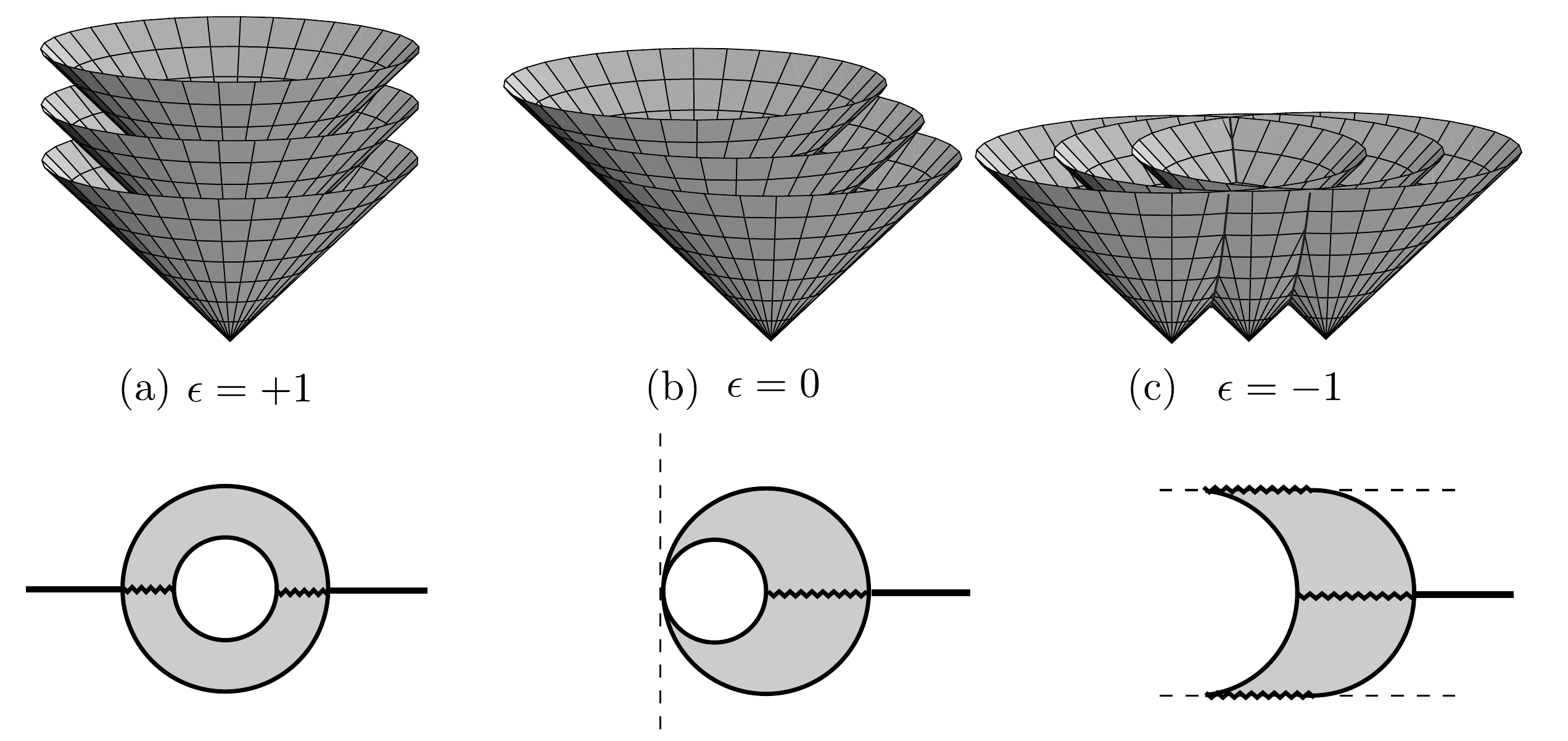}}
%\centerline{\includegraphics[scale=0.55]{allcones}}
\caption{ \small Upper part: The family of future null cones $C_\tau$
with vertices along a timelike (${\epsilon=+1}$), null (${\epsilon=0}$),
or a spacelike (${\epsilon=-1}$) line foliate in three different
ways Minkowski space in the Robinson--Trautman form. Analogous foliations apply
to (anti-)de Sitter space, see \cite{GP:09}.
Lower part: sandwich gravitational waves at a fixed time
for different values of~$\epsilon$ are indicated by the shaded
regions. All wavefronts ${u =\>}$const.\ are spherical (hemispherical for
${\epsilon=-1}$) and expand with the speed of light. The inner region ${u<0}$
is free of topological defects, while the external region ${u>u_1>0}$ contains a
cosmic string (thick line) which ``disintegrates'' within the sandwich (zigzag
line), generating the wave.
}
\label{allcones}
\end{figure}

The simplest choice is to consider \emph{special geodesic trajectories} with
velocity normalised to $-\epsilon$, i.e., a static timelike observer
(${\epsilon=+1}$), a null geodesic (${\epsilon=0}$), or a spacelike
(tachyonic with infinite speed) geodesic (${\epsilon=-1}$). For these choices the
hypersurfaces $C_\tau$ are shown in Figure~\ref{allcones}.
It can be seen that the most natural choice is ${\epsilon=+1}$ which
gives  a (global) foliation of the spacetime. The cones nicely fit one into
another and the wave surfaces at any time form concentric spheres. It is thus
the best candidate for performing the impulsive limit of sandwich gravitational
waves, resulting in the impulse located on a \emph{single} wavefront ${u=0}$.

The family of such sandwich waves was introduced  by Griffiths and Docherty
in \cite{GriffithsDocherty:2002} and further studied in
\cite{GriffithsDochertyPodolsky:2004} for all possible values of $\Lambda$ and
signs of $\epsilon$. The metric has the Robinson--Trautman canonical form
(\ref{RTNmetric}), (\ref{RTP}), with  the function $F(\zeta,u)$ taken to be
 \begin{equation}
 F(\zeta,u)= \zeta^{g(u)},
 \label{RTF}
 \end{equation}
where $g(u)$ is \emph{any positive function} of the retarded time $u$.
Consequently,
\begin{equation}
  P^2=\frac{\left[1+\epsilon
(\zeta\bar\zeta)^g\right]^2}{g^2(\zeta\bar\zeta)^{g-1}}\,, \qquad
  (\log P)_{,u} =-\left( 1 +{\textstyle\frac{1}{2}}
\log(\zeta\bar\zeta)^g
\left[ \frac{1-\epsilon (\zeta\bar\zeta)^g}{1+\epsilon (\zeta\bar\zeta)^g}
\right]  \right)\frac{\>g'}{g} ,
   \label{functions}
\end{equation}
where ${g'=g_{,u}}$, and
${\Psi_4=-\frac{1}{2}(1/r)(\zeta/\bar\zeta)\big(|\zeta|^{-g}+\epsilon |\zeta|^{g}\big)^2(g'/g)\,}$
is the only non-trivial component of the Weyl tensor. The solution is thus \emph{conformally flat} (i.e., Minkowski or 
(anti-)de Sitter background) if and only if $g$ is a \emph{constant}. %(${g'=0}$).
In general, this is an exact Robinson--Trautman gravitational wave with an arbitrary profile determined by $g(u)$. Interestingly, the term $(\log P)_{,u}$ in (\ref{RTNmetric})
and also $\Psi_4$ are \emph{both} proportional to the
same wave profile, namely $g'/g$.

The simplest sandwich-wave is obtained using the continuous function $g(u)$
given by
\begin{equation}
  g(u)=1      \hbox{\quad for\ } u<0\,,\quad
  g(u)=1-a\,u   \hbox{\quad for\ } u\in[0,u_1]\,,\quad
  g(u)=1-a\,u_1 \hbox{\quad for\ } u>u_1\,,
   \label{simplest_g}
\end{equation}
where ${a, u_1}$ are positive constants \cite{GriffithsDocherty:2002},
\cite[Sec.\ 19.2.3]{GP:09} so that ${g'=-a}$ within $[0,u_1]$. Outside
this wavezone ${g'=0}$ so that the spacetime is (conformally) flat. However,
ahead of this sandwich wave in the region ${u>u_1>0}$ there is a topological
defect at ${\zeta=0}$ or ${\zeta=\infty}$ (since ${g<1}$) representing a
\emph{cosmic string} with the deficit angle ${2\pi au_1}$. The region ${u<0}$
behind the wave contains no such defect (because ${g=1}$). The solution
(\ref{RTF}), (\ref{simplest_g}) has thus been interpreted as a \emph{breaking
of a cosmic string} in a conformally flat background in which the tension of
the string (deficit angle) reduces uniformly to zero. Such a cosmic string
decay generates a gravitational wave, see the lower part of
Figure~\ref{allcones}.

The derivative of the function $g$ given by (\ref{simplest_g}) has
discontinuities ${|g'|=a}$ at ${u=0}$ and ${u=u_1}$, so that there are
\emph{shocks} on the initial and final wave surfaces of the sandwich (for
discussion of spherical shocks see \cite{Nutku}). More
general families of sandwich waves without such discontinuities in the metric
and $\Psi_4$ can be constructed by considering \emph{smooth} functions
$g(u)$. Moreover, if ${g=1}$ on \emph{both sides} of the sandwich,
the Minkowski or (anti-)de~Sitter background \emph{does not contain a cosmic
string} either in front of nor behind the wave.
\smallskip

\noindent
{\bf Robinson--Trautman impulses. }
Using the model (\ref{simplest_g}), it is easy to perform the \emph{impulsive
limit} of such Robinson--Trautman sandwich waves by taking the limit ${u_1\to
0}$, keeping ${\gamma\equiv au_1>0}$ fixed. This yields the function
\cite{PodolskyGriffiths2004}
\begin{equation}
   g(u)=1        \hbox{\quad for\ } u<0\,,\qquad
   g(u)=1-\gamma \hbox{\quad for\ } u>0\,,
   \label{simplest_g_impulse}
\end{equation}
i.e., ${g(u)=\exp[\,c\,\Theta(u)]}$, where ${c=\ln(1-\gamma)<0}$ and $\Theta$ is
the step function, in which case
\begin{equation}
  g'/g = c\,\delta(u)\,,
   \label{deltaprofile}
\end{equation}
where $\delta(u)$ is the Dirac delta. We thus indeed obtain an \emph{impulsive
gravitational wave}, with the Weyl curvature tensor %(\ref{Psi4case})
localised on the single wavesurface ${u=0}$. Notice that the
Dirac $\delta$ also directly enters the metric via the  ${(\log
P)_{,u}}$ term in (\ref{functions}). This leads us out of the
Geroch--Traschen class \cite{GT:87} of metrics, but due to the simple geometrical structure
it is still possible to calculate the curvature as a distribution.
The ansatz (\ref{RTF}) has also been generalised to obtain sandwich and impulsive
waves with a richer structure, for example two impulses or a bending string, see \cite{GriffithsDochertyPodolsky:2004}.

Alternatively, the family of Robinson--Trautman type~N metrics (\ref{RTNmetric})
can also be expressed in terms of Garc\'{\i}a--Pleba\'nski coordinates
\cite{GDP,SaGaPl:1983} as
  \begin{equation}
\d s^2 =  2\,(r/\psi)^2\,\big|\d\xi-f\,\d u\big|^2
  +2\,\d u\,\d r -\left(2\epsilon - r\,Q-{\textstyle\frac{1}{3}}\Lambda
r^2\right)\d u^2 \,,
  \label{RTNGDP}
  \end{equation}
where ${\psi=1+\epsilon\xi\bar\xi}$,  $\,f(\xi,u)$ is an arbitrary complex
valued function, holomorphic in $\xi$, and %$Q$ is defined as
${Q=(f_{,\xi}+\bar f_{,\bar\xi})-2\epsilon\psi^{-1}(\bar\xi f+\xi\bar f)}$, see also
\cite{Nutku,RT1962}. This line element is related to (\ref{RTNmetric}) via
the transformation $\xi\equiv F(\zeta,u)$ with $F_{,u}=f(F(\zeta,u),u)$,
see \cite{[A1],[B8]}.  With the specific choice (\ref{RTF}) convenient
for sandwich and impulsive waves, this corresponds to
\cite{PodolskyGriffiths2004}
  \begin{equation}
  f(\xi,u)=(g'/g)\,\xi\log\xi\,.  \label{f}
  \end{equation}
%(This metric is listed in \cite{SaGaPl:1983} as the Robinson--Trautman solutions
%with axial symmetry.)
In particular, for (\ref{simplest_g_impulse}),
(\ref{deltaprofile}) which represents a snapping string accompanied by an
impulsive spherical gravitational wave localised at ${u=0}$, the functions are
  \begin{equation}\label{irt}
  f = c\,\xi\log\xi\,\delta(u)\,,\qquad
  Q = 2c\,\Big( 1+\frac{1-\epsilon\xi\bar\xi}{1+\epsilon\xi\bar\xi}\log|\xi| \Big)\delta(u)\,,
  \end{equation}
see \cite{PodolskyGriffiths2004}. However, observe that the form of the metric
(\ref{RTNGDP}) with (\ref{irt}) has to be considered as being only formal,
since it is quadratic in $f$ and so explicitly contains a \emph{square} of the Dirac~$\delta$. %delta.

Finally, the Robinson--Trautman impulsive spacetimes (\ref{RTNGDP}) with ${f
\equiv f(\xi)\,\delta(u)}$ can also be rewritten in the alternative form
\begin{equation}
\d s^2 = \frac{ 2\,(v/\psi)^2 \,|\d\xi-f(\xi)\delta(\bar u)\,\d \bar u |^2
+2\,\d \bar u\,\d v
  - 2\epsilon\,\d \bar u^2  - v\,Q(\xi) \delta(\bar u)\,\d \bar u^2}
{[\,1+\frac{1}{6}\Lambda \bar u(v-\epsilon \bar u)\,]^2}  \,,
  \label{modif}
\end{equation}
where the profile function ${f(\xi)}$ is \emph{any complex valued (holomorphic)
function} of the spatial coordinates $\xi$. This is obtained from the
Garc\'{\i}a--Pleba\'nski coordinates (\ref{RTNGDP}) by the transformation
\begin{equation}
  r = \frac{v}{1+\frac{1}{6}\Lambda \bar u(v-\epsilon \bar u)}\,,\qquad
  u = \int\frac{\d \bar u}{1-\frac{1}{6}\Lambda \epsilon \bar
u^2}\quad\hbox{such that}\quad u=0\Leftrightarrow \bar u=0\,,
\end{equation}
implying ${\delta(u)\d u=\delta(\bar u)\d \bar u}$, with the spherical impulse
again located on the null surface  ${\bar u=0}$. Clearly, for any ${\bar
u\not=0}$ the metric (\ref{modif}) is the Minkowski or (anti-)de~Sitter
background in the form (\ref{U<0}) with the %trivial
identification ${\bar u=U}$, ${v=V}$, ${\xi=Z}$ (so that ${\psi=p}$).
However, it also has to be considered as only formal since again a \emph{square} of the Dirac
delta enters the metric.

The relation of (\ref{modif}) to the continuous metric form (\ref{en0}) for
expanding impulsive waves was found in \cite{[B8]}. Performing the discontinuous
transformation ${x_i= X_i+\Theta(U)\,[ X_i^{{\hbox{\tiny inv}}}(X_j)-X_i\,]}$,
where ${x_i\equiv(\bar u,v,\xi)}$, ${X_i\equiv(U,V,Z)}$, and $X_i^{{\hbox{\tiny
inv}}}(X_j)$ are specific functions obtained by composing the inverse of
(\ref{inv}) with (\ref{transe3}), it is possible to put (\ref{modif}) formally
into the continuous form (\ref{en0}). This relation is obvious for ${U\not=0}$
using the identity  ${\bar u(v-\epsilon \bar u)= U(V-\epsilon U)}$ (trivially
valid for ${U<0}$ since ${ x_i= X_i}$, and also for ${U>0}$ where ${ x_i=
X_i^{{\hbox{\tiny inv}}}(X_j)}$ since ${\eta\bar\eta-\U\V = U(V-\epsilon U)}$
due to the first identity in (\ref{identitiesABCDEF})). Keeping the
distributional terms arising from $\Theta$ and its derivative in the
transformation, we \emph{formally} obtain also the impulsive terms proportional
to $\delta$ and $\delta^2$ in (\ref{modif}) with
${f(\xi)\equiv Z^{{\hbox{\tiny inv}}}(U=0)-Z}$. Of course, much technical work
is still required before such a discontinuous transformation can be put into a
mathematically sound context.

\subsection{ Impulses generated by infinitely accelerating
sources}\label{sec:2.4}

Specific expanding spherical impulses can also be obtained from \emph{exact
solutions for accelerating sources} in the limit of
\emph{unbounded acceleration}. It was realised by Bi\v{c}\'ak and
Schmidt \cite{BiSc89,Bi90} and corroborated by Podolsk\'{y} and
Griffiths \cite{[B11],[B12]} that such impulses can be obtained from the
family of boost-rotation symmetric solutions
\cite{BicSch89b,PravdaPravdova:2000} which describe the gravitational field of
uniformly accelerating objects, typically attached to conical singularities.
\smallskip

\noindent
{\bf Limit of the Bonnor--Swaminarayan and related solutions.}
Of particular interest is a special case of the Bonnor--Swaminarayan
solution \cite{BonSwa64,BonGrifMacCal94} described by Bi\v{c}\'ak, Hoenselaers
and Schmidt in \cite{BHS1,BHS2} which represents \emph{two particles} of the
Curzon--Chazy type
accelerating in opposite directions. In the limit of infinite
acceleration such a metric can be written as
\begin{equation}
 \d s^2= \textstyle{\frac{1}{4}}(\tilde v-\tilde u)^2 \,e^{-\mu} \,\d\phi^2
 + \textstyle{\frac{1}{4}}(\tilde v+\tilde u)^2 \,e^\mu \,\d \chi^2 -e^\lambda
\,\d \tilde u\,\d \tilde v \,,
 \label{BSnull}
\end{equation}
where ${\mu= -M=\>}$const., ${\lambda=\left[\,\Theta(\tilde u \tilde
v)-\Theta(-\tilde u \tilde v)\,\right]M}$
for the two semi-infinite \emph{receding cosmic strings}
located along the axis ${\rho\equiv\frac{1}{2}(\tilde u-\tilde v)}{=0}$.
The metric is only locally bounded with $\lambda$ suffering a finite jump of
$2M$ on the null cone ${\,\tilde u\tilde v=0}$ which again brings us out of the Geroch--Traschen
class \cite{GT:87}. The resulting spacetime is locally flat
except on the expanding sphere which is the impulsive gravitational wave, generated by two null
particles which move apart in the flat background and are connected to infinity
by two semi-infinite strings.

It is possible to perform a transformation to coordinates in
which the metric is Lipschitz continuous \cite{[B11]}.
It actually brings (\ref{BSnull}) exactly in the form of Gleiser and Pullin \cite{GlePul89}
constructed via their ``cut and paste'' method. Moreover, as shown in \cite{[B11]}, this metric
can be cast in the classic form
(\ref{en0}) with a \emph{real constant} function $H$,
for which, however, the geometric interpretation in terms of the
stereographic correspondence is more obscured.

The above construction has been extended in \cite{[B11]} to a much larger class of
boost-rota\-tio\-nally symmetric spacetimes allowing to attribute an \emph{arbitrary multipole structure} to the receding particles \cite{BHS2}, which, however, vanishes in the
impulsive limit.
\smallskip

\noindent
{\bf Infinitely accelerating black holes. }
In the subsequent paper \cite{[B12]} Podolsk\'y and Griffiths also investigated
null limits of another well-know class of solutions with boost-rotation
symmetry, namely the
\emph{C-metric}. As shown in 1970 by Kinnersley and Walker
\cite{KinnersleyWalker:1970},  such a metric represents a pair of
\emph{uniformly accelerating  black holes}, each of mass $m$.
Their acceleration $A$ is caused either by a strut between the black holes or by
two semi-infinite strings connecting them to infinity. In \cite{[B12]} the limit
 ${A\to\infty}$ was investigated, demonstrating that (scaling $m$ to zero such
that ${mA=\>}$const.) it is again \emph{identical} to the metric
of a spherical  impulsive gravitational wave generated either by a snapping
string, or an expanding strut.

It was natural to expect that the analogous null limit of infinite acceleration
${A\to\infty}$ of a more general C-metric with a \emph{cosmological
constant} ${\Lambda}$ (see \cite{PlebanskiDemianski:1976,[B13],[C1],[C2]}),
would generate an expanding spherical impulsive wave (\ref{en0})
in the (anti-)de~Sitter universe. Such limit turned out to be mathematically
more involved but was finally performed in  \cite{PodolskyGriffiths2004} using
the Robinson--Trautman form extending (\ref{RTNmetric}) to type~D spacetimes. The limit yielded exactly the impulsive metric form with ${ F(\zeta,u)= \zeta^{g(u)}}$ where
${g(u)=\exp[\,c\,\Theta(u)]}$ and $c$ is determined by $mA$, that is
(\ref{RTF}), (\ref{simplest_g_impulse}) and (\ref{deltaprofile}).

%\subsection{Summary of the construction methods}

%Various methods of construction of expanding impulsive waves, necessarily
%spherical, in spaces of constant curvature and the corresponding references are
%summarised in the following table:

%\bigskip
%\begin{tabular}{|l||c|c|}
%\hline
%   method of construction & ${\Lambda=0}$   & ${\Lambda\not=0}$ \\
%\hline
%\hline
%  ``cut and paste'' & \cite{Pen72}, \cite{NutPen92},
%  \cite{Hogan93}, \cite{Hogan94}, \cite{[B10]}, \cite{GlePul89} &
%  \cite{Hogan92}, \cite{[B8]}, \cite{[B10]} \\
%\hline
%  continuous coordinates   & \cite{NutPen92}, \cite{Hogan93},
%  \cite{Hogan94}, \cite{[B8]}, \cite{[B10]}, \cite{NutAli} & \cite{Hogan92},
% \cite{[B8]}, \cite{[B10]} \\
%\hline
% limits of sandwich RT waves   &  \cite{Pen72}, \cite{Nutku}, \cite{[B8]},
%\cite{[B9]} & \cite{[B8]} \\
% \hline
%  infinitely accelerating sources &  \cite{BiSc89}, \cite{Bi90}, \cite{[B11]},
%\cite{[B12]}
%  & \cite{PodolskyGriffiths2004} \\
%\hline
%\end{tabular}
%\bigskip
%\noindent

\smallskip
Further details on the various construction methods, related topics and references can be found in the reviews \cite{[B15],BarHog2003,GP:09}.

\section{Geodesics in expanding impulsive waves}\label{sec:geo}

In this section we focus on geodesics in expanding spherical impulsive waves
propagating in background spacetimes of constant curvature. Thereby we will
exclusively use the continuous form (\ref{en0}) of the metric. Also previous
work on geodesics in these geometries was solely concerned with this form of
the metric. Note that this is in contrast to \emph{nonexpanding} impulsive waves where the distributional forms of the metric have also widely been used, see \cite[Sec.\ 3.1]{PSSS2015} for a brief overview as well as the recent work \cite{SSLP2015}. The reason is that in the expanding case the distributional forms of the metric (\ref{RTNmetric}), (\ref{RTNGDP}), and \eqref{modif} are more complicated than
those in the nonexpanding case and that (\ref{RTNGDP}) and \eqref{modif}, in
addition, contain much wilder singularities.

Indeed, the explicit form of the geodesics in Minkowski spacetime with
expanding spherical gravitational impulses were presented
in \cite{PodolskySteinbauer2003} using the metric (\ref{en0}) with
${\Lambda=0}$. As indicated in the introduction the method employed was a matching
procedure where the geodesics of the background on either side of the impulse were pasted together in a $C^1$-manner, i.e., by equating the corresponding positions and velocities at
the time of interaction with the impulse at ${U=0}$. Strictly speaking, this procedure
 is mathematically justified only in the case of the
geometrically privileged family of geodesics with ${Z=\mbox{const.}}$ while in
the general case it was \emph{assumed without proof} that the geodesics indeed
are $C^1$-curves. In \cite{PodolskySvarc2010} this procedure was
 generalised to the ${\Lambda\not=0}$ cases. To again employ
the ``$C^1$-matching'' procedure it  had to be assumed that \emph{all}
geodesics crossing the impulsive wave actually are continuously differentiable
curves. With this assumption, in all cases ${\Lambda>0}$, ${\Lambda=0}$, ${\Lambda<0}$ the general results on the explicit form of the geodesics have been obtained and employed for a physical discussion of geodesic motion in specific impulsive solutions, such as
the refraction of geodesics caused by the spherical impulse generated by a snapping cosmic string, i.e., (\ref{en0}) with (\ref{h1}).

It is the main aim of this article to \emph{prove} that the ``$C^1$-matching'' procedure is actually a mathematically valid technique. This, in particular, includes an argument that the geodesics are indeed curves of regularity $C^1$, but actually more is needed (cf.\ \cite[Remark 4.1]{PSSS2015}). In fact, we have to prove the following facts on the geodesics in the impulsive wave spacetimes:
\begin{itemize}
\item the geodesics heading towards the impulse \emph{cross} it,
\item they are \emph{unique}, and
\item they are \emph{continuously differentiable}, i.e., of $C^1$-regularity.
\end{itemize}
It is only under these circumstances that
the matching of the geodesics of the background spacetimes --- by equating their
positions and velocities at the instant of interaction with the impulsive wave
--- is guaranteed to give the correct answer.

\subsection{The geodesic equations}
\label{ssec:geo}

We will start out by explicitly deriving the geodesic equations in the real
version of the continuous metric \eqref{en0} which will also enable us to perform a detailed analysis of the form and the regularity of the resulting system of ordinary differential
equations. We consider the metric in the form  (\ref{en0}):
 \begin{equation}
\d s^2 =  \frac{2\left| (V/p)\,\d Z+U_+\,p\,\bar H\,\d\bar Z \right|^2+2\,\d
U\,\d V -2\epsilon\,\d U^2}
{\big[\,1+\frac{1}{6}\Lambda U(V-\epsilon U)\,\big]^2}\,,
 \label{eq:eiwc}
 \end{equation}
where ${p=1+\epsilon Z\bar Z}$, ${\epsilon=-1,0,+1,}$ and again
${H(Z)=\frac{1}{2}[{h'''}/{h'}-({3}/{2})({h''}/{h'})^2]}$
is the Schwar\-zian derivative of an arbitrary complex function $h(Z)$. However,
it will be more convenient to work with the real form of (\ref{eq:eiwc}) which we obtain by setting
${Z=\frac{1}{\sqrt2}(X+\hbox{i}\,Y)}$, namely
\begin{align} \nonumber
\d s^2\ = &\frac{1}{[1+\frac{1}{6}\Lambda U(V-\epsilon U)]^2}\
\bigg( \Big[\frac{V^2}{p^2}+\Up^2\,p^2\,|H|^2\Big](\d X^2+\d Y^2)
\\
&\qquad  + 2\,\Up V\Big[\,\Re(H)\,(\d X^2-\d Y^2) -2\,\Im(H)\,\d X\d Y\Big]
+2\,\d U \d V-2\epsilon \,\d U^2\bigg)\,,
\end{align}
which we will write as
\begin{equation}\label{eq:m}
\d s^2=\frac{1}{\omega^2(U,V)}\left(g_{ij}(U,V,X^k)\,\d X^i\d X^j+2\,\d U\d V-2\epsilon\,
\d U^2\right)\,,
\end{equation}
where
\begin{equation}\label{eq:omega}
\omega = 1+{\textstyle \frac{1}{6}}\Lambda U(V-\epsilon U)  \,.
\end{equation}
Here $\Re(H)$ and $\Im(H)$ denote the real and imaginary parts of the complex
valued function $H$, respectively, and $X^i=(X,Y)$, ${i=1,2}$. The components of $g_{ij}$ are explicitly given by
\begin{align}
 g_{11}&=V^2/p^2+\Up^2\,p^2|H|^2+2\,\Up V\,\Re(H)\,, \label{g1}\\
 g_{22}&=V^2/p^2+\Up^2\,p^2|H|^2-2\,\Up V\,\Re(H)\,, \label{g2}\\
 g_{12}&=-2\,\Up V\,\Im(H)\,. \label{g3}
\end{align}
Observe that --- apart from singularities of $p H$ --- the first two components $g_{11}$, $g_{22}$ contain (in that order)
a smooth term, a term which is $C^{1,1}$ (its first derivative is Lipschitz continuous), and a Lipschitz
continuous term, denoted as $C^{0,1}$. The term $g_{12}$ is just Lipschitz
continuous. Moreover, these three Lipschitz continuous terms in (\ref{g1})--(\ref{g3}) are the only ingredients \emph{of
critical regularity, i.e., below $C^{1,1}$}.
Recall that by Rademacher's theorem, (locally) Lipschitz
continuous functions are differentiable almost everywhere with derivative
belonging (locally) to $L^\infty$. Derivatives of the metric coefficients
$\Up$, $\Up^2$ will always be understood in this sense.

The non-trivial contravariant components corresponding to the metric (\ref{eq:m}) are
\begin{equation}\label{eq:contrametric}
g^{UV}=\omega^2\,,\quad
g^{VV}=2\epsilon\, \omega^2\,,\quad
g^{UU}=0\,,\quad\hbox{and}\quad
\omega^2 g^{ij} \,,
\end{equation}
where $g^{ij}$ is the inverse matrix to $g_{ij}$.
The only non-zero Christoffel symbols of (\ref{eq:m}) are:
\begin{align}
 &
 \Gamma^U_{\,UU}=-\frac{2}{\omega}(\omega_{,U}+\epsilon\omega_{,V})\,,\quad
 \Gamma^U_{\,ij}=\frac{\omega_{,V}}{\omega}\,g_{ij}-\frac{1}{2}g_{ij,V}\,,\\
 &
 \Gamma^V_{\,VV}=-2\frac{\omega_{,V}}{\omega}\,,\quad
 \Gamma^V_{\,VU}=2\epsilon\frac{\omega_{,V}}{\omega}\,,\quad
 \Gamma^V_{\,UU}=-\frac{2\epsilon}{\omega}(\omega_{,U}+2\epsilon \omega_{,V})\,,\\
 &
 \Gamma^V_{\,ij}=\frac{g_{ij}}{\omega}(\omega_{,U}+2\epsilon\omega_{,V})-(\epsilon
g_{ij,V}+\frac{1}{2}g_{ij,U})\,,\\
 &
 \Gamma^{i}_{\,Vj}=-\delta^{i}_{j}\frac{\omega_{,V}}{\omega}+\frac{1}{2}g^{il}g_{jl
,V}\,,\quad
 \Gamma^i_{\,Uj}=-\delta^i_j\frac{\omega_{,U}}{\omega}+\frac{1}{2}g^{il}g_{lj,U}\,,\quad
 \Gamma^{i}_{\,jk}=\,^{(s)}\Gamma^{i}_{\,jk}\,.
\end{align}
Here $\,^{(s)}\Gamma^{i}_{\,jk}$ denotes the Christoffel symbols of the ``spatial
metric'' $g_{ij}$, and
\begin{equation}\label{eq:omegadiff}
\omega_{,V} = {\textstyle \frac{1}{6}}\Lambda U  \,,\qquad
\omega_{,U} = {\textstyle \frac{1}{6}}\Lambda (V-2\epsilon U)  \,.
\end{equation}
Observe that $\,^{(s)}\Gamma^{i}_{\,jk}$, $\Gamma^U_{\,ij}$, and
$\Gamma^i_{\,Vj}$, are Lipschitz continuous, while the Christoffel symbols
containing a $U$-derivative of $g_{ij}$, namely $\Gamma^V_{\,ij}$ and
$\Gamma^i_{\,Uj}$, are merely $L^\infty_{\mbox{\scriptsize loc}}$. All other
Christoffel symbols are at least Lipschitz continuous, hence not of critical regularity.

The geodesic equations thus take the following explicit form
\begin{align}
 \label{geq2}
 \ddot U&
   -\frac{2}{\omega}(\omega_{,U}+\epsilon\omega_{,V})\,\dot U^2
   +\Big(\frac{\omega_{,V}}{\omega}\,g_{ij}-\frac{1}{2}g_{ij,V}\Big)\dot
X^i\dot X^j=0\,,\\
\label{geq1}
 \ddot V&
  -2\frac{\omega_{,V}}{\omega}\,\dot V^2+
  4\epsilon\frac{\omega_{,V}}{\omega}\,\dot V\dot U
   -\frac{2\epsilon}{\omega}(\omega_{,U}+2\epsilon \omega_{,V})\dot U^2\nonumber\\
   &+\Big(\frac{g_{ij}}{\omega}(\omega_{,U}+2\epsilon\omega_{,V})-(\epsilon
   g_{ij,V}+\frac{1}{2}g_{ij,U})\Big) \dot X^i\dot X^j=0\,,\\
 \label{geq3}
 \ddot X^i&
   -\Big(2\delta^{i}_{j}\frac{\omega_{,V}}{\omega}-g^{il}g_{jl,V}\Big)\dot V\dot X^j
   -\Big(2\delta^{i}_{j}\frac{\omega_{,U}}{\omega}-g^{il}g_{jl,U}
\Big)\dot U\dot X^j +\,^{(s)}\Gamma^{i}_{\,jk}\dot X^j\dot X^k=0\,.
\end{align}
In terms of regularity, observe that all of the above equations contain
Lipschitz continuous terms, which, from the perspective of classical
ODE-theory, pose no problem at all. However, the $V$- and the $X$-equations
in addition contain the $L^\infty_{\mbox{\scriptsize loc}}$-terms $g_{ij,U}$, which force
us to go beyond classical existence theory for ODEs. Also observe that the
system is ``fully coupled''  --- in contrast to the nonexpanding case and {\it pp\,}-waves in particular --- so that we cannot decouple either of the equations from the rest of
the system.

To apply the Filippov solution theory in the following subsection, we need to rewrite the geodesic equations \eqref{geq2}--\eqref{geq3} in first order form. Thus, the resulting system is
\begin{align}
 \dot U =&\  \tilde{U}\,,\qquad
 \dot V = \tilde{V}\,,\qquad
 \dot X^i = \tilde{X}^i\,,
   \label{firstorder1}\\
 \dot{\tilde{U}} =&\   \frac{2}{\omega}(\omega_{,U}+\epsilon\omega_{,V})\,\tilde{U}^2
   -\Big(\frac{\omega_{,V}}{\omega}\,g_{ij}-\frac{1}{2}g_{ij,V}\Big)\tilde{X}^i\tilde{X}^j\,,
   \label{firstorder2}\\
 \dot{\tilde{V}} =&\  2\frac{\omega_{,V}}{\omega}\,\tilde{V}^2-
  4\epsilon\frac{\omega_{,V}}{\omega}\,\tilde{V}\tilde{U}
   +\frac{2\epsilon}{\omega}(\omega_{,U}+2\epsilon \omega_{,V})\tilde{U}^2 \nonumber\\
   &-\Big(\frac{g_{ij}}{\omega}(\omega_{,U}+2\epsilon\omega_{,V})-(\epsilon
   g_{ij,V}+\frac{1}{2}g_{ij,U})\Big) \tilde{X}^i\tilde{X}^j\,,
   \label{firstorder3}\\
 \dot{\tilde{X}}^i =&\   \Big(2\delta^{i}_{j}\frac{\omega_{,V}}{\omega}-g^{il}g_{jl,V}\Big)\tilde{V}\tilde{X}^j
 +\Big(2\delta^{i}_{j}\frac{\omega_{,U}}{\omega}-g^{il}g_{jl,U}
\Big)\tilde{U}\tilde{X}^j -\,^{(s)}\Gamma^{i}_{\,jk}\tilde{X}^j\tilde{X}^k\,,
  \label{firstorder4}
\end{align}
which is now a first-order system with discontinuous right hand side.

\subsection{Existence of $C^1$-geodesics}\label{sec:ex}
Geodesic equations with discontinuous right hand side have recently been
solved in the class of nonexpanding impulsive gravitational waves by going
beyond classical ODE-theory. More precisely, in \cite{LSS:13} the
geodesics in impulsive {\it pp\,}-waves have been treated using Carath\'eodory's
solution concept (see e.g. \cite[Ch.\ 1]{F:88}), using the fact that there the
$U$-equation decouples from the rest of the system. In the case of nonexpanding
impulsive waves with non-vanishing $\Lambda$, the $U$-equation is coupled to
the spatial equations, which made it necessary to go even beyond
Carath\'eodory theory. In fact, employing the more general Filippov solution
concept \cite[Ch.\ 2]{F:88} in \cite{PSSS2015} we were able to prove existence,
uniqueness, and $C^1$-regularity of the geodesics in \emph{all} nonexpanding
impulsive gravitational waves on constant curvature backgrounds, thus justifying the previous use of the ``$C^1$-matching procedure'' in these geometries.

Given the fact that in the present case the geodesic equations \eqref{firstorder1}--\eqref{firstorder4} are all coupled together, we will also employ the \emph{Filippov solution concept}. For a short review we refer to
\cite{C:08}, and for the present context to \cite[Appendix]{PSSS2015}. The key
idea is to replace the discontinuous right hand side ${F\colon
\mathbb {R}^d\supseteq D\rightarrow \mathbb{R}^d}$ of a first order system of ODEs
\begin{equation}\label{eq-ode}
 \dot z(t) = F(z(t))\qquad (t\mbox{\ in some interval\ } I)\,,
\end{equation}
by the \emph{set-valued function} defined as
\begin{equation*}
 \mathcal{F}[F](z)\equiv\bigcap_{\delta>0}\bigcap_{\mu(S)=0}
\mathrm{co}(F(B_\delta(z)\backslash S))\,,
\end{equation*}
where $\mathrm{co}(A)$ denotes the closed convex hull of a set $A$ (i.e., the
intersection of all closed and convex supersets of $A$),
$B_\delta(z)$ is the closed Euclidean ball around $z$ of radius
$\delta$, and $\mu$ is the Lebesgue measure.
Hence $\mathcal{F}[F]$, the Filippov set valued map associated with $F$,
averages the values of $F$ in a
neighbourhood of a point $z$ of discontinuity in the following precise sense:
$\mathcal{F}[F](z)$ is given as the intersection of convex hulls of the images
under $F$ of shrinking closed balls around $z$, while ignoring sets $S$ of measure
zero. Clearly at points ${z\in D}$ where $F$ is continuous the set
$\mathcal{F}[F](z)$ is the singleton $\{F(z)\}$, hence if $F$ is continuous 
everywhere, the classical theory is recovered.

Finally, a \emph{Filippov solution} of \eqref{eq-ode} on an interval
$[a,b]\subseteq I$ is an absolutely continuous curve $z:[a,b]\rightarrow D$,
that satisfies
the \emph{differential inclusion}
\begin{equation}\label{eq-di}
 \dot z(t) \in \mathcal{F}[F](z(t))
\end{equation}
almost everywhere. Recall that a curve ${z:[a,b]\rightarrow \mathbb{R}^d}$ is
said to be \emph{absolutely continuous} if for every ${\varepsilon>0}$ there is
a ${\delta>0}$ such that for all collections of non-overlapping intervals
${([a_i,b_i])_{i=1}^m}$ in ${[a,b]}$ with ${\sum_{i=1}^m(b_i-a_i)<\delta}$ we
have that ${\sum_{i=1}^m \|z(b_i)-z(a_i)\|< \varepsilon}$. Moreover, recall that
an absolutely continuous curve is continuous and differentiable almost
everywhere.

Of course, if on a subdomain of $D$ the right hand side $F$ is continuous,
any Filippov solution is also a classical $C^1$-solution of \eqref{eq-ode} there. However, Filippov solutions exist under much more general conditions.
In particular, the question of existence and regularity of the geodesics
follows from a general result for locally Lipschitz continuous semi-Riemannian
metrics, as in \cite{PSSS2015}:

\begin{thm}[Theorem 2 in \cite{S:14}]\label{thm-ex}
Let $(M,g)$ be a smooth manifold with a $C^{0,1}$-semi-Rie\-mann\-ian metric
$g$. Then there exist Filippov solutions of the geodesic equations which are
$C^1$-curves.
\end{thm}

This immediately translates to our setting (\ref{eq:m}) to yield:

\begin{cor}[Existence]\label{cor:ex}
 For the entire class of expanding impulsive gravitational waves on any
 background of constant curvature described by the continuous form of the
 metric (\ref{en0}) with smooth $H$ we have: Given a point ${\cal P}$ and any direction
 $v\in T_{\cal P} M$ there exists a solution in the sense of Filippov to the geodesic equation
 with this initial data, which is a $C^1$-curve.
\end{cor}

The regularity of the geodesics is actually slightly better. Their velocity is
even absolutely continuous, a fact which we will also use in the next
subsection.

\begin{rem}[Local existence for non-smooth $H$]\leavevmode
In physical models of expanding impulses the function $H$ may have singularities. For example, in the case of a spherical impulse generated by a snapped cosmic string (\ref{h1}), described by ${ H=\frac{1}{2}\delta(1-\frac{1}{2}\delta)\,Z^{-2}}$, there is a pole at ${Z=0}$ corresponding to the location of the string. However, in general we still have local existence of geodesics in any region where $H$ is sufficiently smooth (for any ${Z\not=0}$ in the above case).
\end{rem}

Hence we are provided with the existence of $C^1$-geodesics, and we now turn to the
more subtle issues of uniqueness and the fate of geodesics reaching
the wave impulse.

\subsection{Uniqueness of geodesics and crossing of the impulse}\label{sec:uni}

Observe that uniqueness of geodesics is lost in \emph{general} locally Lipschitz
spacetimes. In fact, the threshold for unique (even classical) solvability of the
geodesic equation is the regularity class of $C^{1,1}$-metrics. If one lowers
the regularity only slightly below $C^{1,1}$, e.g.\ by considering metrics in
any  H\"older class $C^{1,\alpha}$ with ${\alpha<1}$, classical counterexamples
(in the Riemannian case)
due to \cite{HW,Hartman} not only show the failure of uniqueness but also of the
usual local convexity properties. However, in the present case the metric
\emph{in addition} to being locally Lipschitz is also \emph{smooth off a null
hypersurface}. In particular, it is piecewise smooth and uniqueness of geodesics
only becomes an issue at points on the wave impulse.
Indeed, uniqueness of the geodesics can be established combining results
from \cite[Section 2.10, p.\ 106]{F:88} with geometric arguments.

In fact, recently we have applied this approach to investigate geodesics in spacetimes with  \emph{nonexpanding} impulses
\cite{PSSS2015}. The core of this argument rested on the fact that the null hypersurface which supports
the impulse is totally geodesic. This is clearly not true in the present case of \emph{expanding} impulses since
${\mathcal{N}=\{U=0\}}$ is a null cone. Complementarily, the methods to be
employed here could not have been used for the
nonexpanding case. To be more precise, the main reasons allowing for
a direct geometric approach are:
\begin{itemize}
 \item the terms in the $U$-equation \eqref{geq2} are continuous (as opposed to
the nonexpanding case), and
 \item in case of geodesic velocities tangent to $\mathcal{N}$, we exploit the
fact that the essentials of the geometry of null hypersurfaces are also valid
in $C^{0,1}$-spacetimes.
\end{itemize}

First, let us elaborate on the first point above. Let $\gamma=(U,V,X^i)$ be a geodesic given by Corollary \ref{cor:ex} and
recall that $\gamma$ is $C^1$. Thus, since $\omega$ is smooth, $g_{ij}$ and $g_{ij,V}$ are (Lipschitz) continuous and
$\dot U$, $\dot X^i$ are continuous, we see that the terms in the $U$-equation \eqref{geq2} are continuous. Consequently,
the component of the Filippov set-valued map corresponding to the $U$-equation is just singleton-valued (see
subsection~\ref{sec:ex}) and so $U$ satisfies \eqref{geq2} almost everywhere.
Moreover, $\dot U$ is \emph{absolutely} continuous so $\ddot U$ satisfies
\eqref{geq2} everywhere, thus $U$ is~$C^2$.

Second, observe that even for a continuous metric it is true that vectors tangent to
a null hypersurface $\mathcal{N}$, with null normal vector field $L$ (i.e.,
${T_{\cal P}\,\mathcal{N}=L_{\cal P}^\perp}$), are either null and proportional to $L$, or
spacelike. Actually the classical argument carries over verbatim.
Moreover, in a locally Lipschitz spacetime we have that the  Levi-Civita
connection satisfies the metric property (i.e., ${\nabla g=0}$ almost everywhere,
see \cite{LM:07,S:08}) and hence again the standard argument applies to show
that the integral curves of $L$ are geodesics that generate $\mathcal{N}$.
Consequently, in our case, we may call $\partial_V$ the null generator of
$\mathcal{N}$. However, any geodesic starting at a null cone in the direction
of a spacelike tangent vector immediately leaves the null cone. In our case
this is manifestly seen from the fact that at  $\mathcal{N}$ the equation for
$U$ takes the form
\begin{align}\label{udd}
 \ddot
U&-\frac{\Lambda}{3}V\dot U^2 - \frac{V}{p^2}(\dot X^2+\dot Y^2)=0\,,
\end{align}
using that $U$ is $C^2$. Because we have ${V>0}$ globally, this allows for
trivial solutions ${U=0}$ (and thus ${\dot U=0=\ddot U}$) only if
${\dot\gamma(0)=\dot V\partial_V+\dot X\partial_X+\dot Y\partial_Y}$ is
proportional to $\partial_V$, the null generator of $\mathcal{N}$ (otherwise ${\dot X^2+\dot Y^2\not=0}$, violating
(\ref{udd})).

With these preliminary observations, our strategy is now to directly use results
of \cite[Sec.\ 10.2]{F:88} and combine them with geometric arguments.  To fix
notations, assume
that ${D\subseteq \mathbb{R}^d}$ is connected and separated by a
smooth hypersurface $N$ into two domains $D^+$ and $D^-$. Let $F$ and
$\frac{\partial F}{\partial x^i}$, ${i=1,\ldots,d}$, be continuous in $D^+$ and $D^-$.
Denote by $F^+$ (respectively $F^-$) the extensions of
$F|_{D^+}$ (respectively $F|_{D^-}$) to the
boundary $N$ and write
$F^+_n$ and $F^-_n$ %, $h_n$ be
for the projections of $F^+,F^-$ onto
the normal to $N$ directed from $D^-$ to $D^+$ at the points of $N$.
Now we have:

\begin{lem}[Sufficient conditions for uniqueness, Lemma 2.10.2 in \cite{F:88}]\label{lem:uniq}
If for ${z_0\in N}$ we have ${F^+_n(z_0)>0}$, then in the domain $D^+$ there exists a unique
Filippov solution of \eqref{eq-ode} starting at~$z_0$. Analogous assertions hold for $D^-$ and
${F^-_n(z_0)<0}$.
\end{lem}

We now translate our problem into the language of the above result. To this end we rewrite the geodesic equations (\ref{geq2})--(\ref{geq3}) in first order form \eqref{firstorder1}--\eqref{firstorder4} in the $8$ variables
\begin{align}
 z=(z^1,\dots,z^8)=(x^\mu,\dot x^\mu)=(U,V,X^i,\tilde{U},\tilde{V}, \tilde{X}^i)\,,
\end{align}
($\mu=0,\dots,3$
and $i=1,2$) with the equation taking the form
\begin{align}
\dot z\ =\ F(z)=\big(\tilde{U}, \tilde{V},\tilde{X}^i,-\Gamma^\mu_{\alpha\beta}(x^\gamma)\,\dot
x^\alpha\dot x^\beta\big)\,.
\end{align}
Since the impulse is located at the null hypersurface $\mathcal{N}=\{U=0\}$
we define $\mathcal{D}^{+}:=\{U>0\}$ to be the ``outside'' of the null cone,
and set $\mathcal{D}^{-}:=\{U<0\}$ to be its ``inside''. Analogously we define
$N\subseteq\mathbb{R}^8$ by $N:=\{z\in\mathbb{R}^8:\
z^1\equiv x^0\equiv U=0\}$ and
set $D^+:=\{z\in\mathbb{R}^8:\ z^1>0\}$ and $D^-:=\{z\in\mathbb{R}^8:\ z^1<0\}$.
Then the first standard unit vector $e^1\in\mathbb{R}^8$ is a normal to $N$
pointing from $D^-$ to $D^+$ and hence $F^+_n=F^-_n=\dot U$.
\medskip

Now consider a geodesic ${\gamma=(U,V,X^i):\ [0,T)\to M}$ of Corollary
\ref{cor:ex} which starts at a point ${\gamma(0)={\cal P}\in\mathcal{D}^+}$ ``outside'' the
null cone, i.e., with ${U(0)=U_0>0}$ and that meets the impulse located at the
null cone ${\{U=0\}}$ (for the first time) at a parameter value $\tau_i$. With 
these assumptions,
we clearly have ${\dot U(\tau)<0}$ for all
${\tau<\tau_i}$ near enough to $\tau_i$, and so by the $C^1$-property of
the geodesics ${\dot U(\tau_i)\leq 0}$. We now distinguish two cases:
\begin{enumerate}
 \item[(1)] $\gamma$ meets $\mathcal{N}$ \emph{transversally}, i.e.
 ${\dot U(\tau_i)<0}$, and
 \item[(2)]$\gamma$ meets $\mathcal{N}$ \emph{tangentially}, i.e.
 ${\dot U(\tau_i)=0}$.
\end{enumerate}

In the first case, clearly Lemma \ref{lem:uniq} applies to guarantee that
$\gamma$ continues uniquely to negative values of $U$ (i.e., to the ``inside'' of
the null cone $\mathcal{D}^{-}$). Also, this case is ``time symmetric'', that is
if a geodesic starts with negative $U$-values (i.e., ``inside'' the cone, that
is in $\mathcal{D}^-$) and hits $\mathcal{N}$
transversally then it continues uniquely to positive values of $U$, (i.e., to
the ``outside'' $\mathcal{D}^+$).
\medskip

At this point we make the following essential observation: For
geodesics $\gamma$ in the sense of Theorem~\ref{thm-ex} in general
$C^{0,1}$-spacetimes the scalar product of their tangent
$g(\dot\gamma,\dot\gamma)$ and hence their causal character is \emph{not necessarily
preserved:} Indeed, the usual argument fails since $\dot\gamma$ only needs to obey
an inclusion relation (at almost all points) rather than
${\nabla_{\dot\gamma}\dot\gamma=0}$.

Returning to our case and to a geodesic $\gamma$ of Corollary \ref{cor:ex} starting in
$\mathcal{D}^+$ we clearly have that
${g(\dot\gamma(\tau),\dot\gamma(\tau))=g(\dot\gamma(0),\dot\gamma(0))=:c}$ as long
as $\gamma$ stays in $\mathcal{D}^+$, i.e., for ${\tau<\tau_i}$, since there it
satisfies the smooth geodesic equation. Moreover, by the $C^1$-property we have
that also ${g(\dot\gamma(\tau_i),\dot\gamma(\tau_i))=c}$. However, unless we know
that $\gamma$ only hits $\mathcal{N}$ in isolated points we \emph{cannot} infer
that $g(\dot\gamma,\dot\gamma)=c$ globally, since, in principle, $\gamma$ could
stay for some time within the wave surface $\mathcal{N}$ and there its derivative again would only satisfy the inclusion relation (almost everywhere). We will, however,
\emph{prove} in the course of our discussion that this does not happen
and that \emph{all} geodesics $\gamma$ starting in $\mathcal{D}^+$ (resp.\ in
$\mathcal{D}^-$) and hitting $\mathcal{N}$ only do so in isolated points and
hence the scalar product of their tangent as well as their causal character is globally preserved.
\medskip

In fact, we have already (almost) established this for case (1),
i.e., for all geodesics $\gamma$ hitting $\mathcal{N}$ transversally either
from the ``outside'' or from the ``inside''. By the above, all such $\gamma$
uniquely continue immediately to the ``inside'' (resp.\ to the ``outside'') and
hence $g(\dot\gamma,\dot\gamma)$ is preserved. This completely settles the case
for all \emph{causal} (i.e., timelike or null) geodesics of case (1) since they
cannot hit the null cone $\mathcal{N}$ twice.  All such $\gamma$ are
\emph{globally unique} solutions of  the respective initial value problem and
moreover they meet $\mathcal{N}$ at the \emph{single instant} $\tau_i$ of
(parameter) time which finally implies that $g(\dot\gamma,\dot\gamma) $ is
globally preserved.

We are left with $\gamma$ of case (1) starting out \emph{spacelike} in
$\mathcal{D}^+$ and hitting $\mathcal{N}$. Again $\gamma$ enters the interior
$\mathcal{D^-}$ immediately with unchanged $g(\dot\gamma,\dot\gamma)$, hence
stays spacelike and will eventually hit $\mathcal{N}$ again. By an
argument given below it actually again hits transversally. Then
by our above discussion of the ``time symmetric'' case, $\gamma$ again crosses
$\mathcal{N}$ uniquely and proceeds in a spacelike manner back to the
``outside''. So such $\gamma$  again are \emph{globally unique} solutions of
the respective initial value problem meeting $\mathcal{N}$ at
\emph{two isolated instants} of (parameter) time and finally
$g(\dot\gamma,\dot\gamma) $ is globally preserved. This now completely settles
case (1).
\medskip

Turning to case (2), we first note that the scalar product of the geodesic
tangent upon hitting the impulse satisfies
\begin{align}
 g(\dot\gamma,\dot\gamma)\Big|_{\tau=\tau_i}=
  \Big(\,2\dot U(\dot V-\epsilon \dot U)
   +\frac{V^2}{p^2}(\dot X^2+\dot Y^2)\,\Big)\Big|_{\tau=\tau_i}\,,
\end{align}
which in case (2) (that is, $\gamma$ hits $\mathcal{N}$ tangentially) further
simplifies to
\begin{align}\label{eq:tnorm}
   g(\dot\gamma,\dot\gamma)\Big|_{\tau=\tau_i}\, = \frac {V^2}{p^2}(\dot
X^2+\dot Y^2)\Big|_{\tau=\tau_i}\geq 0 \,,
\end{align}
since in this case ${\dot U(\tau_i)=0}$. By the above discussion this is
impossible for all geodesics $\gamma$ that started out timelike in
$\mathcal{D}^+$ (resp.\ $\mathcal{D}^-$). Hence for such $\gamma$ we find
ourselves exclusively in case (1) which we have already settled: hence we are
done with \emph{all timelike} geodesics.

To deal with case (2) we thus only need to consider geodesics that start out
either null or spacelike in $\mathcal{D}^+$ and we will distinguish two subcases:
\begin{enumerate}
 \item [(2a)] $\dot\gamma(\tau_i)$ is \emph{null},
 \item [(2b)]  $\dot\gamma(\tau_i)$ is \emph{spacelike}.
\end{enumerate}

In case (2a), $\dot \gamma(\tau_i)$ is actually proportional to the null
generator $\partial_V$ of $\mathcal{N}$ and if we were in a smooth spacetime
we could conclude immediately by uniqueness that $\gamma$ could not have started
in $\mathcal{D}^+$ in the first place. In our situation we have to be more careful and argue as follows: The geodesic $\gamma$ for ${\tau<\tau_i}$ lies in $\mathcal{D}^+$ hence satisfies
the smooth geodesic equation and consequently, by the
$C^1$-property, ${0=g(\dot\gamma(\tau_i),\dot\gamma(\tau_i))=g(\dot\gamma(\tau),
\dot\gamma(\tau))}$ for all ${\tau<\tau_i}$. So $\gamma$ is a null geodesic also
for ${\tau<\tau_i}$, that is in $\mathcal{D}^+$, and hence also a (smooth) null geodesic in the \emph{background} spacetime, which by assumption hits $\mathcal{N}$ tangentially.
By the continuity of its tangent, it also hits $\mathcal{N}$ tangentially in the background spacetime, which is clearly not possible. So such a $\gamma$ does not exist, and similarly in the ``time symmetric'' case there does not exist any null geodesic $\gamma$ starting in the ``inside'' $\mathcal{D}^-$ of the null cone and hitting $\mathcal{N}$ tangentially at ${\tau=\tau_i}$ with $\gamma(\tau_i)$ being a null vector.
\medskip

This argument also proves the fact that if a geodesic ${\gamma=(U,V,X^i)}$
in the sense of Corollary~\ref{cor:ex} at some parameter value $\tau_0$
satisfies ${U(\tau_0)=0}$ and $\dot\gamma(\tau_0)$ is
proportional to the generator $\partial_V$ of $\mathcal{N}$, then $\gamma$ lies
entirely in $\mathcal{N}$. Then it even follows that $\gamma$ is one of the null
generators: its velocity being tangent to $\mathcal{N}$ is either null and in
$\mathrm{span}(\partial_V)$ or spacelike. The latter possibility is ruled out
since it would cause $\gamma$ to leave $\mathcal{N}$ and consequently the
$V$-equation \eqref{geq1} reduces to $\ddot V=0$.

Thus also the solutions of the geodesic equation with data
$\gamma(0)\in\mathcal{N}$ and $\dot\gamma(0)$ \emph{null} are unique.
Moreover the argument to be laid out in the following paragraph also establishes
this fact for spacelike $\dot\gamma(0)$. Observe that it is the geometry that
leads to this conclusion, which seems rather unexpected just from looking at the
equations which in this case are merely differential inclusions.
\medskip

Finally, we are left with discussing case (2b), where we know already that
$\gamma$ started out in $\mathcal{D^+}$ as a \emph{spacelike} geodesic. Using
the conditions ${U(\tau_i)=0=\dot U(\tau_i)}$
the geodesic equation (\ref{geq2}) implies
\begin{align}
 \ddot U(\tau_i)=\frac{V}{p^2}\big(\dot X^2+\dot Y^2\big)\Big|_{\tau=\tau_i}\> >\> 0\,,
\end{align}
where positivity follows from condition (2b) inserted into (\ref{eq:tnorm})
and keeping in mind that we have ${V>0}$ anyway.
Hence $U$ has a strict local minimum at $\tau_i$, and consequently $\gamma$ which
started in $\mathcal{D}^+$
with positive values of $U$ returns to positive $U$-values, hence touches $\mathcal{N}$ just
at the single instant $\tau_i$ and continues uniquely into $\mathcal{D}^+$ as
a spacelike geodesic. In particular, it stays outside the null cone and actually
is a geodesics of the background outside the impulse, hence smooth. To end
this discussion, observe that here no ``time symmetric'' case exists, since a
geodesic starting with negative $U$-values, i.e., in $\mathcal{D}^-$ cannot
attain ${U(\tau_i)=0}$ and at the same time have a minimum at $\tau_i$. Hence this
excludes the existence of geodesics spacelike in the ``inside'' and hitting
$\mathcal{N}$ tangentially, a fact which we have already used above.
\medskip

Summing up, we have proved that all geodesics of Corollary~\ref{cor:ex}
are unique solutions of the respective initial value problem. Moreover we have
gained complete information on their behaviour when meeting the impulse:

\begin{thm}[Uniqueness]\label{thm:uni}
 For the entire class of expanding impulsive gravitational waves on any
 background of constant curvature described by the continuous form of the
 metric (\ref{en0}) with smooth $H$ we have: Given any point ${\cal P}$ and any
 direction ${v\in T_{\cal P} M}$  there exists a \emph{unique} $C^1$-solution $\gamma$
in the sense of Filippov to the geodesic equations with this initial data.

Moreover, if such a geodesic meets the impulsive wave located at
${\mathcal{N}=\{U=0\}}$ at all, it is either one of its null generators or it
hits it in \emph{isolated points}.
 \end{thm}

Consequently we globally have:
\begin{cor}[Preservation of causal character]\label{cor:3.5}
The geodesics $\gamma$ of Theorem~\ref{thm:uni} satisfy
 \begin{align}
  g(\dot\gamma,\dot\gamma)=\mbox{constant}\,,
 \end{align}
 and, in particular, the causal character of $\gamma$ can be defined
 globally.
\end{cor}

Another conclusion to be drawn from the above discussion and Theorem~\ref{thm:uni} concerns
the actual behaviour of the geodesics starting off the wave impulse and
hitting it:

\begin{cor}[Crossing the expanding impulse]\label{cor:3.6}
 The geodesics of Theorem~\ref{thm:uni} that start off the wave surface
 $\mathcal{N}=\{U=0\}$ and hit it at all, do so in isolated points either
\begin{itemize}
\item [(a)] \emph{transversally} and pass from the ``outside'' $\mathcal{D}^+$
  to the ``inside'' $\mathcal{D}^-$, or vice versa, or
\item [(b)] \emph{tangentially}, in which case they are spacelike
  and come from the ``outside'' $\mathcal{D}^+$  and revert to the ``outside'' $\mathcal{D}^+$ again.
 \end{itemize}
\end{cor}

\begin{rem}[Uniqueness for non-smooth $H$]\leavevmode
If the function $H$ has singularities then, given arbitrary initial data, the
geodesic equation possesses locally defined unique $C^1$-solutions in any region
where $H$ is sufficiently smooth.
\end{rem}

Finally, we see that all the necessary facts have been established to state our
main achievements on the geodesics in all expanding impulsive gravitational
waves propagating in constant curvature backgrounds with any cosmological
constant $\Lambda$:
\begin{enumerate}
 \item The ``$C^1$-matching procedure'' is a mathematically valid
method to explicitly describe the form of the geodesics that \emph{cross} the
wave impulse (i.e., those of Corollary \ref{cor:3.6}(a)).
 \item We have found geodesics that just \emph{touch} the impulse,
i.e., those of Corollary \ref{cor:3.6}(b), which are \emph{not}
not covered by the ``$C^1$-matching''\footnote{Indeed, if one applies the 
matching to such geodesics it becomes trivial in the sense that one has to 
apply the same transformation \eqref{transe3} twice (instead of \eqref{transe3} 
in the outside and \eqref{inv} in the inside, cf.\ section \ref{sec:explgeod}).
Hence the constants from both sides agree, which is of course in perfect 
agreement with the fact that these geodesics are just (smooth) geodesics of the 
background outside the impulse: They do not ``feel'' the impulse at all.} and 
which are actually geodesics of the background spacetime outside the impulse 
and, in particular, smooth. 
\end{enumerate}

\begin{rem}[Completeness]
 Note that the geodesic completeness depends crucially on the topology ``outside'' the impulse (assuming, as usual, that the ``inside'' is a part of the background spacetime without topological defects like cosmic strings). Therefore, no
general statements can be made about the completeness of the geodesics given by
Theorem \ref{thm:uni}. However, since we proved that geodesics that hit the
impulse either cross it to $\mathcal{D}^-$ or return to $\mathcal{D}^+$, the impulsive wave surface is no obstruction to locally continue the geodesics. Thus the only obstructions can
come from global topological effects.
\end{rem}

\section{Explicit $C^1$-matching of geodesics crossing the impulse}\label{sec:explgeod}

To complete our investigation, in this final section we summarise the main
results on the refraction of geodesics by expanding impulses, as derived
previously in \cite{PodolskySteinbauer2003,PodolskySvarc2010}, that have now
been rigorously justified by the results of Section \ref{sec:geo}.

The idea of such a ``$C^1$-matching procedure'' is based on the fact that the
geodesics crossing the impulsive wave surface $\mathcal{N}$ are uniquely defined
$C^1$-curves in the continuous coordinates (\ref{en0}) hence their positions
and velocities at the instant of interaction are \emph{the same on both sides}
of $\mathcal{N}$.

To directly observe the influence of such an expanding impulse, it is 
beneficial to employ relations (\ref{inv}) and (\ref{transe3}), and to transform 
the explicit components of the interaction position and velocity (denoted by 
the subscript $_i$) of the global $C^1$-geodesics from the continuous 
system~(\ref{en0}) into the coordinate system~(\ref{conf}), naturally 
associated with the background spaces of constant curvature.
We do so separately in the regions \emph{outside} the impulse (${U>0}$, the 
superscript $^+$) using (\ref{transe3}),
and \emph{inside} of it (${U<0}$, the superscript $^-$) using (\ref{inv}). By 
combining these expressions we explicitly relate
the parameters of a geodesic approaching the impulse from the region ${U>0}$ to the unique one describing its continuation
in the region ${U<0}$. For the relation between the positions we thus get
\begin{equation}
\U_i^-=|h'|\frac{|Z_i|^2}{|h|^2}\,\U_i^+ \,, \qquad \V_i^-=|h'|\,\V_i^+ \,, \qquad \eta_i^-=|h'|\frac{Z_i}{h}\,\eta_i^+ \,, \label{ref:positions}
\end{equation}
while the relation between the velocities is
\begin{eqnarray}
&& \dot{\U}_i^- = a_{_\U}\dot{\U}_i^+ +b_{_\U}\dot{\V}_i^+ +\bar{c}_{_\U}\dot{\eta}_i^+ +c_{_\U}\dot{\bar{\eta}}_i^+ , \nonumber \\
&& \dot{\V}_i^- = a_{_\V}\dot{\U}_i^+ +b_{_\V}\dot{\V}_i^+ +\bar{c}_{_\V}\dot{\eta}_i^+ +c_{_\V}\dot{\bar{\eta}}_i^+ , \label{ref:velocities} \\
&& \dot{\eta}_i^- \,= a_{\eta}\,\dot{\U}_i^+ +b_{\eta} \dot{\V}_i^+ +\bar{c}_{\eta}\,\dot{\eta}_i^+ +c_{\eta}\,\dot{\bar{\eta}}_i^+ , \nonumber
\end{eqnarray}
where
\begin{align}
& a_{_\U} = \frac{1}{|h'|}\left|1+\frac{Z_i}{2}\frac{h''}{h'}\right|^2 \,,\qquad  b_{_\U} =  \frac{|h|^2}{|h'|}\left|1+\frac{Z_i}{2}\bigg(\frac{h''}{h'}-2\frac{h'}{h}\bigg) \right|^2 \,, \nonumber \\
& \qquad c_{_\U} = -\frac{h}{|h'|}\bigg[1+\frac{Z_i}{2}\left(\frac{h''}{h'}-2\frac{h'}{h}\right)\bigg]\bigg[1+\frac{\bar{Z}_i}{2}\frac{\bar{h}''}{\bar{h}'}\bigg] \,,  \label{coef:calU}\\
& a_{_\V} = \frac{1}{4|h'|}\left|\frac{h''}{h'}\right|^2 \,, \qquad
b_{_\V} = \frac{|h|^2}{4|h'|}\left|\frac{h''}{h'}-2\frac{h'}{h}\right|^2 \,, \qquad
c_{_\V} = -\frac{h}{4|h'|}\left(\frac{h''}{h'}-2\frac{h'}{h}\right)\frac{\bar{h}''}{\bar{h}'} \,, \label{coef:calV}\\
& a_{\eta}= \frac{1}{2|h'|}\bigg(1 +\frac{Z_i}{2}\frac{h''}{h'}\bigg)\frac{\bar{h}''}{\bar{h}'} \,, \qquad b_{\eta} = \frac{|h|^2}{2|h'|}\bigg[1+\frac{Z_i}{2}\bigg(\frac{h''}{h'}-2\frac{h'}{h}\bigg)\bigg]\bigg(\frac{\bar{h}''}{\bar{h}'}-2\frac{\bar{h}'}{\bar{h}}\bigg) \,,\nonumber\\
& \qquad \bar{c}_{\eta}= -\frac{\bar{h}}{2|h'|}\bigg(1+\frac{Z_i}{2} \frac{h''}{h'}\bigg)\bigg(\frac{\bar{h}''}{\bar{h}'}-2\frac{\bar{h}'}{\bar{h}}\bigg) \,,\qquad c_{\eta}= -\frac{h}{2|h'|}\bigg[1+\frac{Z_i}{2}\left(\frac{h''}{h'}-2\frac{h'}{h}\right) \bigg]\frac{\bar{h}''}{\bar{h}'} \,, \label{coef:eta}
\end{align}
and ${\bar{c}_{_\V} = \overline{c_{_\V}}}$,  ${\bar{c}_{_\U} = \overline{c_{_\U}}}$. In accordance with Corollary
\ref{cor:3.5} the velocities preserve the normalisation, namely ${\dot{\eta}_i^-\dot{\bar{\eta}}_i^-
-\dot{\U}_i^-\dot{\V}_i^- = \dot{\eta}_i^+\dot{\bar{\eta}}_i^+ -\dot{\U}_i^+\dot{\V}_i^+}$.

All the coefficients are just constants which are obtained by evaluating the specific function $h(Z)$ and its derivatives at ${Z=Z_i}$ using ${h(Z_i)=\eta_i^+/\V_i^+}$, see \eqref{xi}. Interestingly, these refraction formulas do not depend on the curvature parameter $\epsilon$. Naturally, in the trivial case ${h(Z)=Z}$, i.e., ${H=0}$, they reduce to the identity, which is consistent with the fact that there is no refraction effect in the absence of an impulse.

\section{Conclusion}\label{sec:final}
By employing the continuous form of the metric and the Filippov solution 
concept, we rigorously proved existence and global uniqueness of 
$C^1$-geodesics crossing expanding impulsive gravitational waves which 
propagate in spaces of constant curvature, that is Minkowski, de~Sitter and 
anti-de~Sitter universes. Thereby we have studied the interaction of free test particles 
with such impulsive waves and we have mathematically justified the ``$C^1$-matching 
procedure'' previously used in the literature to derive the explicit form of 
these geodesics.

This work can be understood as a first step in the long-term project of 
understanding the suspected equivalence between the distributional form 
\eqref{RTNGDP} or \eqref{modif} of the expanding wave metric and its continuous 
form \eqref{en0}. To this end we need to understand the behaviour of the 
geodesics in a very precise manner, since they give the key to the 
`discontinuous coordinate transformation' relating the various forms of the 
metric, cf.\ \cite{KS:99} for the \emph{pp}-wave case. Such discontinuous 
transformations will be subject to further investigations, in order to 
obtain a mathematical sound way of describing this equivalence (probably using 
a non-linear theory of generalised functions).

Another interesting issue would be to study the specific causality properties 
of these physically relevant Lorentzian manifolds of low regularity 
to complement the theoretical investigations of \cite{CG:12}, \cite{Sae:15}.

\section*{Acknowledgement}
JP and R\v{S} were supported by the Albert Einstein Center for Gravitation and 
Astrophysics (Czech Science Foundation 14-37086G) and the project 
UNCE~204020/2012. CS and RS acknowledge the support of FWF grants P25326 and
P28770.

\end{document}